\documentclass{article}

\pdfoutput=1 

\usepackage{arxiv}

\usepackage[utf8]{inputenc} 
\usepackage[T1]{fontenc}    
\usepackage{hyperref}       
\usepackage{url}            
\usepackage{booktabs}       
\usepackage{amsfonts}       
\usepackage{nicefrac}       
\usepackage{microtype}      
\usepackage{lipsum}		
\usepackage{graphicx}
\usepackage{natbib}
\usepackage{doi}

\title{Visualising emergent phenomena at oxide interfaces}

\author{ \href{https://orcid.org/0000-0002-3391-8745}{\includegraphics[scale=0.06]{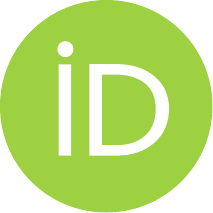}\hspace{1mm}Michael Oberaigner} \\
	Graz University of Technology,\\
	Graz, Austria \\
	\texttt{michael.oberaigner@tugraz.at} \\
	\And
    \href{https://orcid.org/0000-0002-8441-9778}{\includegraphics[scale=0.06]{figure_0.pdf}\hspace{1mm}Manuel Ederer} \\
	Technical University of Vienna,\\
	Vienna, Austria \\
	\texttt{manuel.ederer@tuwien.ac.at} \\
	\And
    \href{https://orcid.org/0000-0002-3689-3336}{\includegraphics[scale=0.06]{figure_0.pdf}\hspace{1mm}Sandeep K. Chaluvadi} \\
	Istituto Officia dei Materiali-CNR,\\
	Trieste, Italy \\
	\texttt{chaluvadi@iom.cnr.it} \\
	\And
    \href{https://orcid.org/0000-0002-1082-9651}{\includegraphics[scale=0.06]{figure_0.pdf}\hspace{1mm}Pasquale Orgiani} \\
	Istituto Officia dei Materiali-CNR,\\
	Trieste, Italy \\
	\texttt{orgiani@iom.cnr.it} \\
	\And
    \href{https://orcid.org/0000-0003-1739-3763}{\includegraphics[scale=0.06]{figure_0.pdf}\hspace{1mm}Regina Ciancio} \\
	Istituto Officia dei Materiali-CNR,\\
	Trieste, Italy \\
	\texttt{ciancio@iom.cnr.it} \\
	\And
    \href{https://orcid.org/0000-0003-0080-2495}{\includegraphics[scale=0.06]{figure_0.pdf}\hspace{1mm}Stefan Löffler} \\
	Technical University of Vienna,\\
	Vienna, Austria \\
	\texttt{stefan.loeffler@tuwien.ac.at} \\
    \And
    \href{https://orcid.org/0000-0002-2116-7761}{\includegraphics[scale=0.06]{figure_0.pdf}\hspace{1mm}Gerald Kothleitner} \\
	Graz University of Technology,\\
	Graz, Austria \\
	\texttt{gerald.kothleitner@felmi-zfe.at} \\
    \And
    \href{https://orcid.org/0000-0003-0755-958X}{\includegraphics[scale=0.06]{figure_0.pdf}\hspace{1mm}Daniel Knez\thanks{Corresponding author}} \\
	Graz University of Technology,\\
	Graz, Austria \\
	\texttt{daniel.knez@felmi-zfe.at} \\
}

\renewcommand{\shorttitle}{2DEG}

\hypersetup{
pdftitle={Visualising emergent phenomena at oxide interfaces},
pdfsubject={cond-mat.mtrl-sci},
pdfauthor={Michael Oberaigner},
pdfkeywords={Transition metal interface, 2DEG, STEM-EELS, DPC, Lanthanum aluminate, Anatase, Oxygen Vacancies, Electronic state mapping},
}

\begin{document}
\maketitle

\begin{abstract}
Knowledge of atomic-level details of structure, chemistry, and electronic states is paramount for a comprehensive understanding of emergent properties at oxide interfaces. We utilise a novel methodology based on atomic-scale electron energy loss spectroscopy (EELS) to spatially map the electronic states tied to the formation of a two-dimensional electron gas (2DEG) at the prototypical non-polar/polar $TiO_2$/$LaAlO_3$ interface. Combined with differential phase contrast analysis we directly visualise the microscopic locations of ions and charge and find that 2DEG states and $Ti^{3+}$ defect states exhibit different spatial distributions. Supported by density functional theory (DFT) and inelastic scattering simulations we examine the role of oxygen vacancies in 2DEG formation. Our work presents a general pathway to directly image emergent phenomena at interfaces using this unique combination of arising microscopy techniques with machine learning assisted data analysis procedures.
\end{abstract}


\section{Introduction}
The electronic, magnetic and catalytic properties of transition metal oxide (TMO) interfaces have received significant attention in recent years. This is due to their promising potential for breakthroughs in materials science and nanotechnology. The formation of an electrically conductive layer is one of the most intriguing phenomena observed in these interfaces. Such two-dimensional electron gases (2DEG) are studied at several TMO interfaces, with the prototypical example being $SrTiO_3$/$LaAlO_3$ forming a non-polar/polar interface~\cite{Ohtomo2004}. Strong electron coupling at such interfaces trigger a metal-insulator transition by different external stimuli, such as electric or magnetic fields, temperature, light and strain \citep{Hwang2012, Yang2011}. Applications of these effects have already been demonstrated, for instance in highly integrated transistors \citep{Cen2009}, or in ultrahigh-density nanoelectronic memory and logic elements~\citep{Park2010}. The metal-insulator transition in such systems is often attributed to an electron transfer at the interface due to a polar discontinuity, referred to as electronic reconstruction or polar catastrophe model~\citep{Nakagawa2006,Ohtomo2004,thiel2006}. Electric field formation, however, can also lead to the formation and accumulation of oxygen vacancies (VO) at the interface region, which act as n-type dopants~\citep{Chen2010, Bristowe2011}. The absence of impact on electron mobility at varying VO concentrations is attributed to the separation of charge carriers and charged oxygen vacancies \citep{Siemons2007}. Moreover, the intermixing of cations may generate charge carriers to establish a conductive region, or it may impede charge mobility as a scatterer within the conductive layer \citep{Schlom2011, Bristowe_2014, Chen2013}.\\
In order to explore the role of these mechanisms a large variety of techniques like angle-resolved photoemission spectroscopy (ARPES), X-ray absorption spectroscopy (XAS), X-ray photoemission spectroscopy (XPS), magnetic/electrical transport measurements or scanning tunneling microscopy and spectroscopy (XSTM/S) have already been applied~\citep{Salluzzo2009, Drear2013, trogila2022, Walker2022, strocov2022dimensionality, thiel2006, Bigi.2020, PhysRevLett.109.246807}. While highly valuable information could certainly be obtained from such measurements, all these techniques either lack spatial resolution or are surface sensitive only. Scanning transmission electron microscopy (STEM) is the only technique that enables atomic imaging of such interfaces in cross-section and, coupled with electron energy loss spectroscopy (EELS), allows to study the spatial distribution of elements, valence states and even individual orbitals~\citep{Knez.2020, maurice2008, LOFFLER201726, Bugnet2022}. Additional integrated differential phase contrast (iDPC) imaging and differential phase contrast (DPC) field mapping enables us to provide a complete picture of the contributing phenomena at unprecedented detail.\\  
By relying on the most recent technological and methodological advancements in electron microscopy, such as direct electron detection and machine learning assisted post-processing, combined with density functional theory (DFT) simulations and inelastic scattering calculations, we reveal the structure, chemistry and electronic states of the $TiO_2$/$LaAlO_3$ interface at atomic scale. Such characterisations are crucial for designing and engineering TMO interfaces with tailored 2DEG properties, which can have far-reaching implications for advanced electronic devices and technologies. We believe that our results will also help to understand similar systems, such as $LaAlO_3$/$LaVO_3$\citep{Wadati.2008} or $SrVO$/$SrTiO$~\citep{Mirjolet.2021}.\\

\begin{figure}[ht]
\centering
\includegraphics[width=16cm]{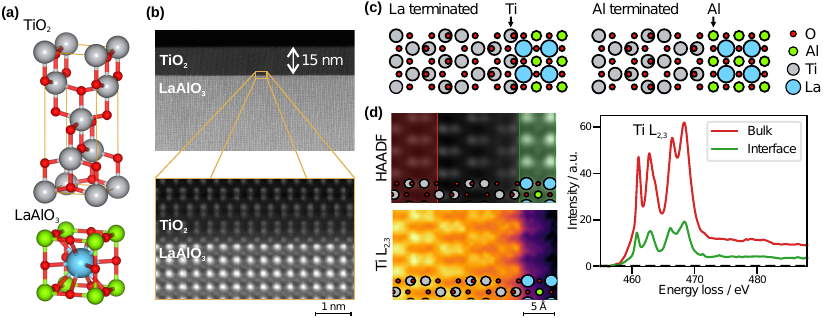}
\caption{$TiO_2$ and $LaAlO_3$ unit cells (a) with HAADF-recorded interface (b) demonstrating the high-quality epitaxial film. The two types of terminations (c) can be determined by HAADF and EELS analysis, integrating the $Ti$ L\textsubscript{2,3} edges (d).}\label{fig:crystal}
\end{figure}

In this work we present a visualisation of the spatial distribution of electronic states associated with the presence of 2DEG and defects at the anatase/LAO ($TiO_2$/$LaAlO_3$) interface (Fig.~\ref{fig:crystal}a). While this interface is closely related to the STO/LAO system, it lacks the influence of interfacial strain due to the almost perfectly matching lattices between anatase and LAO, with only 0.1~\% lattice mismatch compared to the 3~\% mismatch of STO/LAO interfaces \citep{gobaut2017, Cheng.2017, Minohara2014}. Anatase $TiO_2$ (001) is grown epitactic on single crystalline LAO (001) by pulsed laser deposition (PLD). 
Fig.~\ref{fig:crystal}b provides a representative view of the cross-section of the anatase film and interface region, demonstrating that the interface is atomically sharp and of high quality, which has also been shown in previous studies~\citep{Knez.2020,Islam2022}. The anatase film has a thickness of approximately 15~nm. Film and substrate fully match in plane as a consequence of the very low lattice mismatch of anatase with the LAO substrate. The interface can have two types of termination as illustrated in Fig.~\ref{fig:crystal}c, which exhibit different electronic properties. Since the 2DEG is only formed at the La-terminated interface ($Al$-termination leads to semiconducting behaviour), the focus of this study is on La-terminated interfaces \citep{wang2010, Minohara2014, Cheng.2017}. Proof for the presence of the La-termination is provided by high-angle annular dark-field (HAADF) contrast and further supported by EELS mapping of the $Ti$ L\textsubscript{2,3} edge signal, as shown in Fig.~\ref{fig:crystal}d.\\

\section{Electronic Analysis}\label{sec:til}

\begin{figure}[ht]
\centering
\includegraphics[width=16cm]{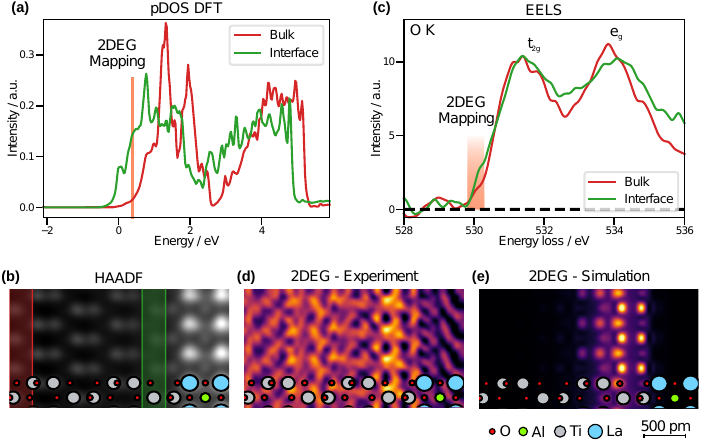}
\caption{2DEG mapping with the O K edge. (a) pDOS calculation showing O atoms with metallic features at the interface. (b) Corresponding HAADF image. (c) EELS-Spectrum comparison of the O K edge between bulk and interface. (d) Experimental map of the 2DEG using the marked regions in (b). (e) Simulated map of the 2DEG using the marked regions in (a).}\label{fig:o_k_2deg}
\end{figure}

The partial density of states (pDOS) calculated by DFT indeed indicates a metallic behaviour for the oxygen layers close to the La-terminated interface due to the crossing electronic states with the Fermi-level and accordingly these states can be attributed to O atoms contributing to the 2DEG formation (Fig.~\ref{fig:o_k_2deg}a). Two regions, one at a distance of approximately 1.5~nm from the interface (denoted by a red rectangle in the corresponding HAADF image (Fig.~\ref{fig:o_k_2deg}b)) is compared with the region at the interface (green rectangle), showing that the metallic features are only present close to the interface. In order to detect the 2DEG at the interface experimentally, we performed monochromated STEM EELS mapping of the O~K edge. By using a post-processing sequence to identify and automatically remove artefact-affected data, as well as correcting sample drift and enhancing signal-to-noise ratio through multi-frame imaging, we are able to detect faint spectral details with high resolution in both space and energy. Details about the applied procedures are provided in the supplementary file. The results of the EELS analysis are summarised in Fig.~\ref{fig:o_k_2deg}c. The features of the O~K edge from the first region exhibits the typical features for anatase, with the t\textsubscript{2g} and e\textsubscript{g} peaks due to the effect of crystal field splitting. A comparison, however, with the region at the interface reveals the presence of an additional feature directly at the edge onset at approximately 530~eV as highlighted in Fig.~\ref{fig:o_k_2deg}c. Such features clearly indicate the occurrence of additional states at the Fermi-level, as seen in the DFT calculations. It shall be noted that for the simulation no VO or cation intermixing was considered, indicating that these states are intrinsic to the interface and not related to VO accumulation. By selecting the corresponding energy region, a map for the spatial distribution of these states within this energy range is generated, as denoted in Fig.~\ref{fig:o_k_2deg}c. The resulting map is shown in Fig.~\ref{fig:o_k_2deg}d. It becomes clear that these 2DEG states are limited to a spatial range between the first two $Ti$ atoms from the interface. For comparison, a similar map is generated from the simulation data by performing STEM-EELS multislice simulations from the pDOS at a specific energy loss (Fig.~\ref{fig:o_k_2deg}e). The simulations match well with the experiment and indicate that the metallic features indeed originate from electronic reconstruction. Despite the delocalised nature of the electrons contributing to the 2DEG, the corresponding states are still localised to the atomic columns due to the nature of EELS projecting onto core states and the channelling by the atomic columns.\\
Similar features at the Fermi-level should also be visible at the $Ti$ L\textsubscript{3} EELS edge, however in this case the signal overlaps with features associated to the presence of defect ($Ti^{3+}$) states \citep{Bigi.2020}, as shown in Supplementary~Fig.~1. Furthermore, multiplet effects, which are not included in DFT calculations \citep{Kruger2010}, affect the shape of Ti L edges and can therefore not be interpreted directly as the pDOS, e.g. pre-peak features at the L\textsubscript{3} edge~\citep{Kucheyev2004}. To include such effects into our calculations would require to solve the Bethe-Salpeter equation, which is computationally expensive and beyond the scope of this work. In contrast, such effects caused by electron-electron interactions can be ignored for the $O$~K edge because the interaction energy of the 1s core electron with valence electrons is less than 1~meV~\citep{Frati2020, GROOT200531}.

\begin{figure}[ht]
\centering
\includegraphics[width=16cm]{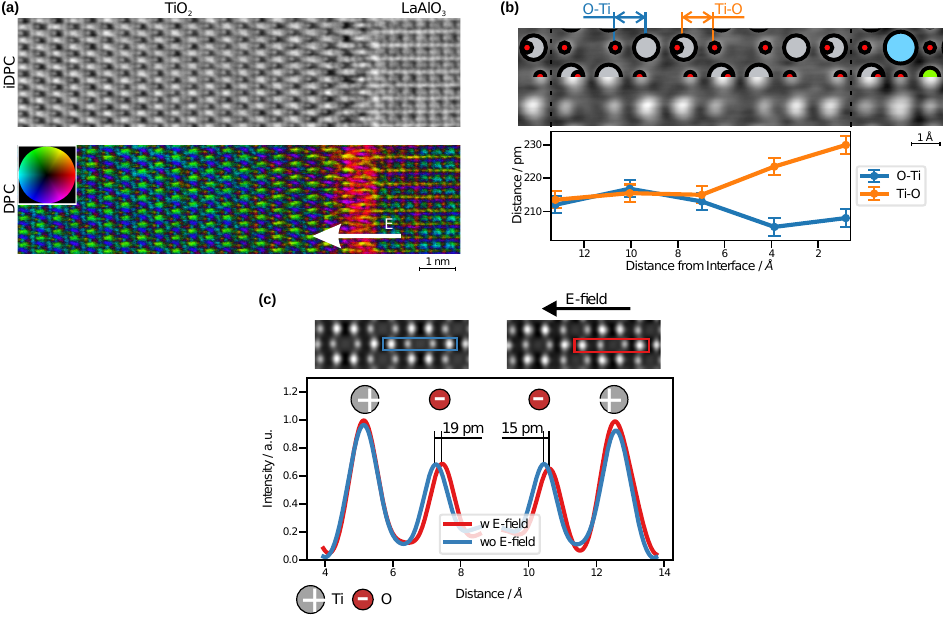}
\caption{DPC experiments at the interface. (a) iDPC and DPC measurements. Colors in the the DPC image correspond to the deflection of the beam with the orientation indicated by the color wheel in the inset. (b) iDPC image showing the stretching and compression of the Ti-O dipoles depending on their orientation in the presence of an electric field. Error bars are estimated by statistic sampling (around 5 measurements per point) and averaged over all points. (c) Simulated iDPC images of anatase with and without electric field demonstrating the atomic displacement in the profile plots. }\label{fig:electric_field}
\end{figure}

The electronic reconstruction model predicts that the positively charged $(LaO)^+$ terminating layer should induce an electric field that compensates for the discontinuity between the polar LAO and the non-polar $TiO_2$, avoiding a polar catastrophe \citep{Ohtomo2004, Pauli_2008}. In order to probe the presence of such local electric field changes associated with the presence of a 2DEG, we performed STEM DPC measurements at the interface region. DPC probes the momentum transfer caused by electromagnetic fields within the specimen to the incident electrons. The results are shown in Fig.~\ref{fig:electric_field}a and indeed reveal an out-of-plane deflection of the electron beam at the interface towards the LAO substrate, which indicates the presence of an confining potential with a field gradient parallel to the interface. The colors correspond to the direction of beam deflection, which means that the field points in the opposite direction. The field region extends approximately~1~nm into the anatase film, which matches with predictions from DFT calculations~\citep{Bigi.2020}.\\ 
To account for the possibility of artefacts due to diffraction contrast variations, which are known to occur particularly at interfaces and grain boundaries~\citep{MACLAREN201557}, additional atomic displacement measurements have been performed. A local electric field is expected to manifest itself by shifts of the negatively charged $O$ ions relative to the positively charged $Ti$ ions. This effect of an electric field on the anatase ion positions is demonstrated by multislice calculations based on molecular dynamics simulations (Fig.~\ref{fig:electric_field}c). These atom displacements are measurable in a high resolution integrated DPC (iDPC) micrograph of the interface region (depicted in Fig.~\ref{fig:electric_field}b). The signal obtained with iDPC has been shown to be particularly efficient for light element imaging~\citep{LAZIC2016265}. From this image the atomic positions are determined with sub-pixel accuracy by fitting 2D Gaussian functions to the columns, utilising the \textit{Atomap} package~\citep{nord2017}. The distances are then measured between pairs of $O$ and $Ti$ atoms in out-of-plane direction. These measurements reveal a stretching and compressing of the lattice close to the interface, dependent on the orientation of the $Ti$-O dipoles relative to the field direction. The distance between the ions changes by approximately 10-20~pm. The evident orientation dependence also allows us to exclude the possibility that these displacements are caused by other effects such as strain or vacancies and clearly indicate the presence of an electric field, with the same out-of-plane direction as found in the DPC field mapping data.

\section{Defects}

\begin{figure}[ht]
\centering
\includegraphics[width=16cm]{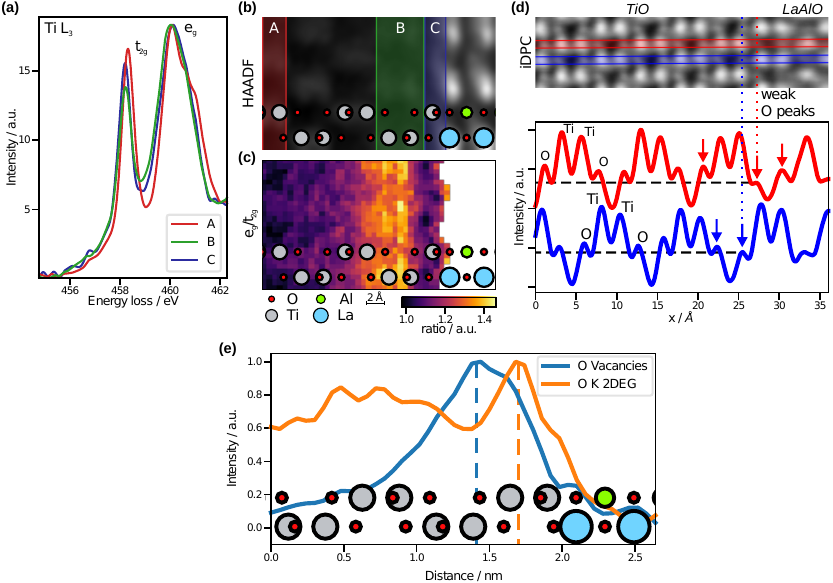}
\caption{VO at the interface. Normalised $Ti$~L\textsubscript{3} edges (a) from regions indicated in (b). (c) Ratio between e\textsubscript{g} and t\textsubscript{2g} from Ti L\textsubscript{3} edges, which have at least 600 counts. (d) iDPC image of the interface with corresponding profiles revealing the lower $O$ peaks. (e) Normalised intensity profile of VO mapped by the $Ti$ e\textsubscript{g}/t\textsubscript{2g} ratio and 2DEG mapped by the $O$ K pre-edge across the interface.}\label{fig:ti_l_edge_ratio}
\end{figure}

Although our 2DEG mapping agrees with the defect-free simulations within the experimental and numerical accuracy, VO are known to play a significant role in influencing the properties of the interface and the formation of the 2DEG~\citep{Kalabukhov2007}. On the one hand VO act as donors, introducing additional electrons and thereby increasing the carrier density at the interface \citep{Siemons2007}. On the other hand, VO also induce lattice distortions~\citep{Knez.2020,Lee.2022}, which affect the electronic structure and band alignment at the interface~\citep{Zhang.2019b}. To probe local variations in the VO concentration the EELS Ti L\textsubscript{3}~e\textsubscript{g}/t\textsubscript{2g} peak ratio and the valley between them are useful indicators. Changes of these features, which are often traced back to changes in the $Ti^{3+}$/$Ti^{4+}$ ratio \citep{maurice2008}, are indeed visible in our EELS measurements as shown in Fig.~\ref{fig:ti_l_edge_ratio}a, where three spectra extracted from regions indicated in (b) are given. To measure the amplitudes Voigt profiles are fitted to the individual peaks (see the Methods section for more details) yielding a map of the e\textsubscript{g}/t\textsubscript{2g} peak amplitude ratio provided in Fig.~\ref{fig:ti_l_edge_ratio}c. This image reveals a higher concentration of $Ti^{3+}$ at a distance up to the second and third layer of $Ti$ from the interface while in contrast, the termination layer of $Ti$ at the interface shows a ratio similar to the bulk, which is in close agreement with the spatial extend of the confining potential and literature reports~\citep{Bigi.2020}. Note that the fitting procedure would result in artefacts and noise in the $LaAlO_3$ regions due to the absence of $Ti$, which is ignored here. Further proof for the presence of a higher amount of VO is provided by iDPC images by a slightly reduced intensity of oxygen columns close to the interface, as shown in Fig.~\ref{fig:ti_l_edge_ratio}d. The terminating oxygen layer at the interface, in contrast, exhibits a higher intensity, which again agrees well with the EELS data. Such accumulation of VO cannot be observed in the LAO (see \textit{Oxygen vacancies in $LaAlO_3$} in supplementary information). \\
The high spatial resolution of our data enables us to separate the location of the VO related defect states from the $O$~K pre-edge states attributed to the 2DEG. Fig.\ref{fig:ti_l_edge_ratio}e shows a comparison of intensity profiles obtained from the experimental maps of the pre-edge states from Fig.~\ref{fig:o_k_2deg}c and the peak ratio e\textsubscript{g}/t\textsubscript{2g} from Fig.\ref{fig:ti_l_edge_ratio}b, corresponding to the respective phenomena. Thereby, a spatial separation of the VO accumulation and the 2DEG is observed. While the 2DEG is found to be located close to the interface, VO are accumulated within a range between approximately 0.5 and 1 nm from the interface.\\ 
Unlike oxygen vacancies, $Al$ and $La$ interdiffusion does not affect the charge carrier density of the 2DEG due to their inefficiency as n-type dopant in $TiO_2$~\citep{Minohara2014}, however cation intermixing can reduce the carrier mobility \citep{Chen2013}. We were able to detect some degree of cation intermixing of $Ti$ and $Al$ between the terminating layers (Supplementary Fig.~3), which will affect the electron mobility.

\section*{Summary}\label{sec12}
In this study, the spatial distributions of the 2DEG, oxygen vacancies, and cation intermixing have been imaged with atomic resolution by STEM-EELS at a non-polar/polar TMO interface. These distributions have a significant impact on the properties of the 2DEG and are therefore important for engineering new TMO interfaces. At the example of $TiO_2$/$LaAlO_3$ interface displayed in this work, the electronic states that cause the 2DEG have been experimentally localised in the first two $Ti$ layers from the termination by STEM EELS. Corresponding inelastic scattering simulations of a defect-free interface match with experimental EELS maps within the experimental and numerical accuracy, indicating that VO are not primarily required for the 2DEG formation, which is in agreement with previous work~\citep{Cheng.2017}. We were further able to disentangle 2DEG states from $Ti^{3+}$ defect states in EELS and showed that VO are accumulated at a distance of around 2-3 $Ti$ layers from the interface by a built-in electric field. We directly visualised this field by DPC field mapping and measured its effects on the positions of the ions within the field. Last, cation interdiffusion of $Al$ in the conductive layer has been observed, which can affect the electron mobility. Our study provides a complete picture off several phenomena accompanying 2DEG formation at oxide interfaces and highlights the great power of combining state-of-the-art STEM experiments, advanced data processing techniques and high-level simulations to address emergent phenomena in complex oxide heterostructures.

\section{Methods}\label{sec13}

\subsection*{Sample preparation}
The thin layer of anatase is epitaxially grown on (001)-oriented $LaAlO_3$ substrates by pulsed laser deposition (PLD) using a $KrF$ excimer pulsed laser source of wavelength 248~nm. A laser energy density of about 2~Jcm\textsuperscript{-2} and a laser repetition rate of 1~Hz were used. The growth temperature was set to 700°C and the target-substrate distance was about 5~cm. A stoichiometric single-crystalline rutile $TiO_2$ target was used and the oxygen pressure was approximately 10\textsuperscript{-4}~mbar during growth. The TEM sample is prepared by focused ion beam milling (FIB) at 30~keV and 5~keV utilising a \textit{Nova 200 NanoLab} from \textit{FEI}. The FIB lamella is polished by low voltage $Ar^+$ ion milling at 900~eV utilising a \textit{Fischione Nanomill}. Before STEM investigations, the specimens are baked at 190°C for 2~h in vacuum and cleaned 10~s with $Ar$/$O$ plasma to reduce contamination. 

\subsection*{STEM-EELS and DPC}
The experiments are performed on an \textit{FEI Titan\textsuperscript{3} G2 60-300} equipped with a monochromator and probe C\textsubscript{s}-corrector. The EELS signal is recorded by a direct electron detection camera, \textit{K2} by \textit{Gatan Inc}. A HAADF image is recorded simultaneously with each spectrum image, which is used as reference image for multi-frame spectrum summation and spatial alignments during post-processing. The images are taken at 300~keV with a EELS semi-collection angle of 24.2~mrad and a beam semi-convergence angle of 15.5~mrad. The beam current was approximately 40~pA with an energy resolution of 250~meV (full width at half maximum of the zero loss peak) by utilising the monochromator. 
For the DPC measurements, the microscope is operated without an excitation of the monochromator. The semi-convergence angle is varied between 15.6~mrad and 19.6~mrad. The signal is recorded by a segmented annular dark field detector, which is divided into four quadrants with a collection angle of 6-34~mrad. The specimen thickness values, which are used for corresponding simulations, are determined by utilising the automatised PACBED analysis based on pre-trained convolutional neural networks as described elsewhere in more detail~\citep{XU201859, oberaigner2023}.

\subsection*{Data post processing}
STEM-EELS data are denoised by a combination of a multi-frame integration approach and multivariate analysis including automatic control mechanisms assessing data quality, like clustering the SI and rating frames at the multi-frame approach by a $\ell_2$-norm. The procedure is described in detail in the supplementary information.\\
To map the $Ti^{+3}$/$Ti^{+4}$ ratio, Voigt profiles, which are a convolution between a Lorentzian and Gaussian function, are fitted into the e\textsubscript{g} and t\textsubscript{2g} peaks from the $Ti$ L\textsubscript{3}~edge. These profiles consider the Lorentzian component, which is caused by the lifetime broadening, and the Gaussian part, which is introduced by the energy spread of the primary electrons. The peak width and the shape factor are determined by fitting into the spectrum of summed SI and these parameters are locked for the amplitude fitting at each pixel afterwards. 

\subsection*{DFT calculation}
All ab-initio density functional theory calculations are performed using the all-electron, full-potential augmented plane wave method as implemented in WIEN2k \citep{blaha2020}. For the exchange and correlation potential, the PBE-GGA \citep{Perdew1996} is chosen. The number of plane waves are limited to 4818 plus 428 local orbitals. The full Brioullin-zone is divided into a $13 \times 13 \times 1$ \textbf{k}-point mesh. For the atomic structure, we have constructed a super-cell consisting of 2.5 anatase unit cells and 4.5 $LAlO_3$ unit cells. The periodic super-cell incorporates a $La$--$Ti$ terminated interface and an $Al$--$Ti$ terminated interface, although only the first is of significance for this work.

\subsection*{Image simulation}
The STEM-EELS images are calculated using inelastic multislice simulations \citep{Loeffler2013phd}. The code incorporates a standard multislice algorithm \citep{kirkland1998} which simulates the elastic electron propagation through the sample. For the inelastic interaction between the probe beam and the sample electrons, the mixed dynamic form factor (MDFF) \citep{Loeffler2013} is calculated. The MDFF includes the atomic site dependent cross density of states and the radial wave functions of the one-electron Kohn-Sham orbitals of the corresponding electronic transition. Both are extracted from the DFT calculation described in the previous section.

\subsection*{Molecular dynamics}
Molecular dynamics (MD) of an anatase supercell (15x15x8 unit cells) are performed with and without an external applied electric field (0.1~\AA) along [001] crystallographic direction. MD simulations were performed with the Large-scale Atomic/Molecular Massively Parallel Simulator (LAMMPS)~\cite{Plimpton.1995} using a modified embedded atom method potential \cite{Joost.2015}. Energy minimisation was achieved by applying a conjugate gradient minimisation with cell size optimisation, followed by a damped dynamics step~\cite{Knez.2020}.

\section*{Data availability}
Extra data is available from the corresponding author upon reasonable request.

\section*{Code availability}
The codes used for data analysis are available from the corresponding authors upon reasonable request.

\section*{Supplementary information}
The Supplementary Information contains more information about

\section*{Acknowledgments}
The authors thank Martina Dienstleder for preparing the FIB-lamella and Johannes Biskupek for additional Nanomill polishing. We thank also Thomas Mairhofer for discussing the DPC experiments and  Alessandro Barla for help with EELS data interpretation.
This project has received fundings from the European Union’s Horizon 2020 Research and Innovation Programme (Grant No. 823717, project “ESTEEM3”) and the Austrian Science Fund (FWF) under grant number I4309-N36 “Orbitalkartierung an Grenzflächen”. For the purpose of open access, the authors have applied a CC BY public copyright license to any Author Accepted Manuscript version arising from this submission. The authors also acknowledge fundings by the Zukunftsfond Steiermark for purchasing the K2 camera and the Wirtschaftskammer Steiermark (WKO Stmk.) for providing additional funds for maintenance.

\section*{Author contributions}
M.O. performed the electron microscopy experiments and data analysis. M.E. performed the DFT calculations and STEM-EELS simulations. S.L, D.K. and G.K. conceived the experiments and supervised the simulations and experiments. D.K performed supporting atomistic simulations. S.K.C and P.O. grew the samples and performed the pre-characterisation of the films. R.C. and P.O. contributed to the discussion of results and to reviewing drafts. M.O. and D.K. wrote the manuscript with the contribution from all the authors.

\section*{Declarations}
The authors declare that they have no competing interests.

\bibliographystyle{unsrtnat}
\bibliography{references}  






\end{document}


\maketitle

\section{Interface}

\subsection{Fine structure mapping by Ti ELNES}
\begin{figure}[ht]\centering
\includegraphics[width=10cm]{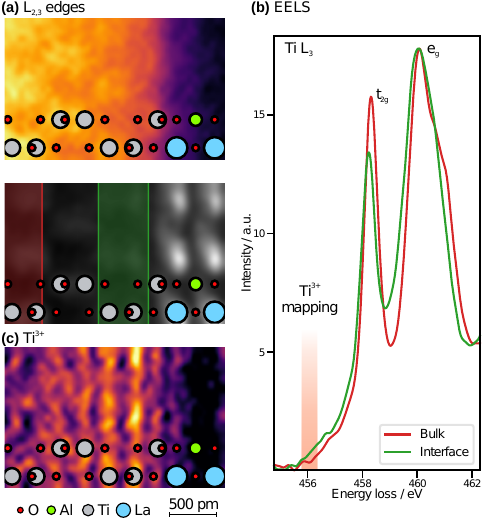}
\caption{Fine structure mapping with the $Ti$ L\textsubscript{3} edge. (a) Mapping of the L\textsubscript{2,3} edge demonstrating $La$-termination with corresponding HAADF image. (b) EELS-Spectrum comparison of the $Ti$ L\textsubscript{3} edge between bulk and interface with the marked integration region to generate the map of $Ti^{+3}$ in (c).}
\label{fig:2deg_ti_mapping}
\end{figure}

The interface can be determined as $La$-terminated by integrating the $Ti$ L edges (Supplementary~Fig.~\ref{fig:2deg_ti_mapping}a). As in reference \citep{Bigi.2020} indicated, the $Ti$ L\textsubscript{3} edge of $Ti^{+3}$ rises earlier than for $Ti^{+4}$. A small integration window at the onset of $Ti$ L\textsubscript{3} edge, which is associated with $Ti^{+3}$, is used to image the distribution of these defects (\ref{fig:2deg_ti_mapping}b). This can be localised to the second and third $Ti$ layer in front of the termination (\ref{fig:2deg_ti_mapping}c). 

\subsection{Cation intermixing}
\begin{figure}[ht]\centering
\includegraphics[width=0.6\linewidth]{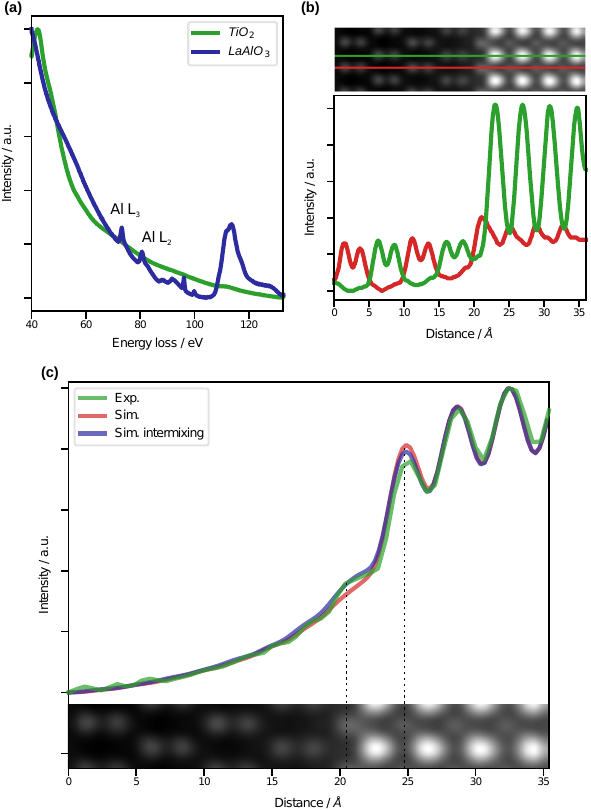}
\caption{Cation intermixing of $Al$ and $Ti$. (a) Recorded EELS low loss region to map $Al$ L\textsubscript{3} edge. (b) HAADF image demonstrating an atomic sharp interface. (c) Profile of $Al$ across the interface.}
\label{fig:elemental_maps}
\end{figure}

The $Al$ L\textsubscript{3} edge is used to determine the $Al$ distribution across the interface (Supplementary~Fig.~\ref{fig:elemental_maps}a). The corresponding HAADF image indicates an atomic sharp interface due to the uniform peak height of all $La$ columns (Supplementary~Fig.~\ref{fig:elemental_maps}b). Supplementary~Fig.~\ref{fig:elemental_maps}c compares the profiles of the $Al$ L\textsubscript{3} edge between experiment and simulation. For the simulation, an atomic sharp interface is used. Both $Al$ L\textsubscript{3} profiles are normalised between 0 and 1. Although the overall agreement between the profiles is good, the experiment indicates some additional $Al$ in the first few layers of $TiO_2$ as well as a lower intensity of the last $Al$ layer. An additional simulation is performed, where 10~\% of the $Ti$ atoms at the terminating $Ti$ layer are exchanged with $Al$ from the terminating $Al$ layer. This model fits the experiment better and suggests some degree of cation intermixing of $Ti$ and $Al$ at the terminating layers. 

\subsection{Oxygen vacancies in $LaAlO_3$}
As with the oxygen vacancy detection in $TiO_2$, the defects are also searched for in $LaAlO_3$. Since the L-edges of $La$ have a very high energy loss at around 6000~eV, the M-edges are investigated. In the Supplementary~Fig.~\ref{fig:o_vacancies_lao}a, the edges of the interface and the bulk are compared. Thereby, neither a shift nor a ratio change of the edges can be observed. Furthermore, the differentiated DPC (dDPC) image does not show any significant changes indicating oxygen defects (Supplementary~Fig.~\ref{fig:o_vacancies_lao}b). Therefore, an accumulation of oxygen vacancies, like at $TiO_2$, can be excluded.

\begin{figure}[ht]\centering
\includegraphics[width=12cm]{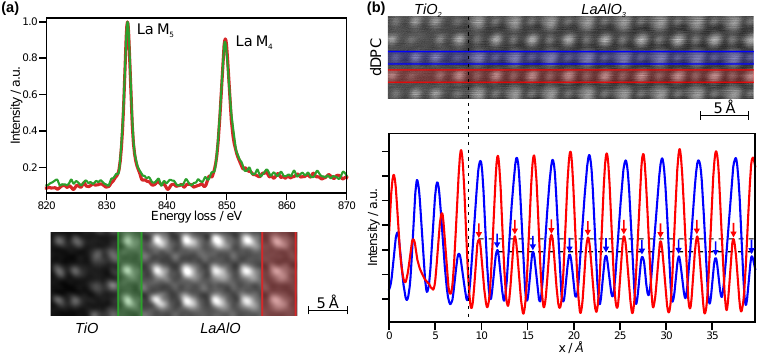}
\caption{Experiments to detect oxygen vacancies in the $LaAlO_3$ towards the interface. (a) Comparison of the M-edges of $La$ in the bulk and at the interface. (b) dDPC image of the interface with corresponding profiles.}
\label{fig:o_vacancies_lao}
\end{figure}

\section{Post processing}

\begin{figure}[ht]\centering
\includegraphics[width=0.5\linewidth]{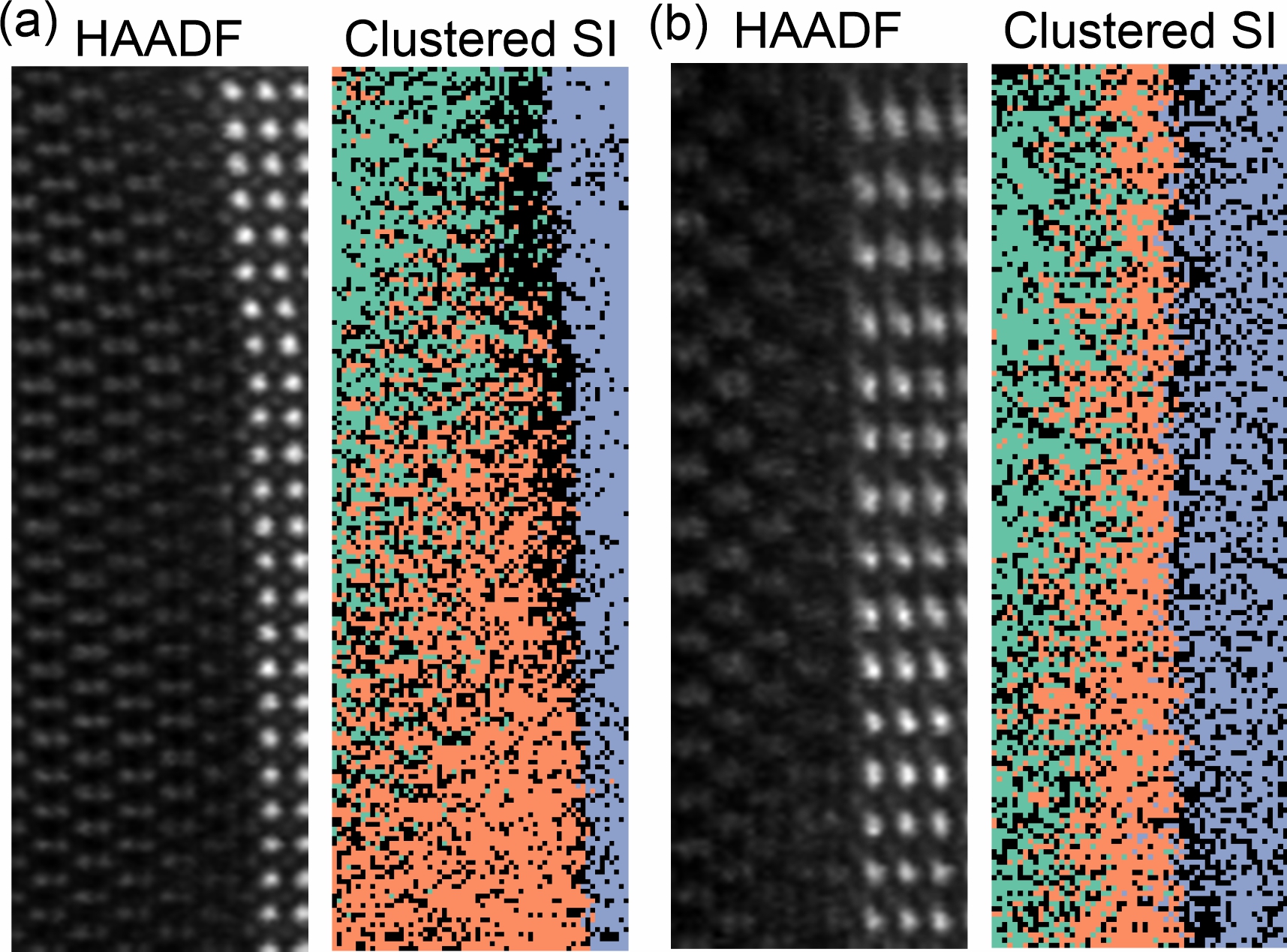}
\caption{Clustering algorithm on two different spectrum images. (a) No uniform interface, due to decreasing beam current along the interface. (b) Uniform interface. For a more detailed analysis, the averaged spectra of each cluster can be investigated.}
\label{fig:clustering}
\end{figure}

Averaging a multi-frame spectrum image without considering that EELS spectra in different areas can have slightly different features, will cause mixing of the signals and ultimately smear out features that are important for mapping orbitals. Therefore, the spectrum image is clustered to detect and visualise changes in the EELS spectra, which can not be seen in the dark field image. For this purpose, an algorithm suggested by \citep{ryu2021} for unsupervised classification of EELS spectra is implemented. In short, the dimension and the noise of the spectrum image is first reduced by the efficient PCA algorithm. To account for the shot noise dominated spectra from the direct detection camera, we use a weighted PCA (wPCA) \citep{keenan2004}. The dimension is further decreased with the non-linear dimensionality reduction method t-SNE (t-distributed Stochastic Neighbor Embedding) \citep{hinton2002}. The signal is then clustered using the unsupervised OPTICS (Ordering Points To Identify the Clustering Structure) algorithm, which does not require a priori knowledge of the number of clusters, making it possible to detect unknown clusters \citep{ankerst1999}. By clustering the spectrum image, slight changes caused by various influences can be visualised, like decreasing beam current (Supplementary~Fig.~\ref{fig:clustering}). These regions can be excluded from further processing.

\begin{figure}[ht]\centering
\includegraphics[width=1\linewidth]{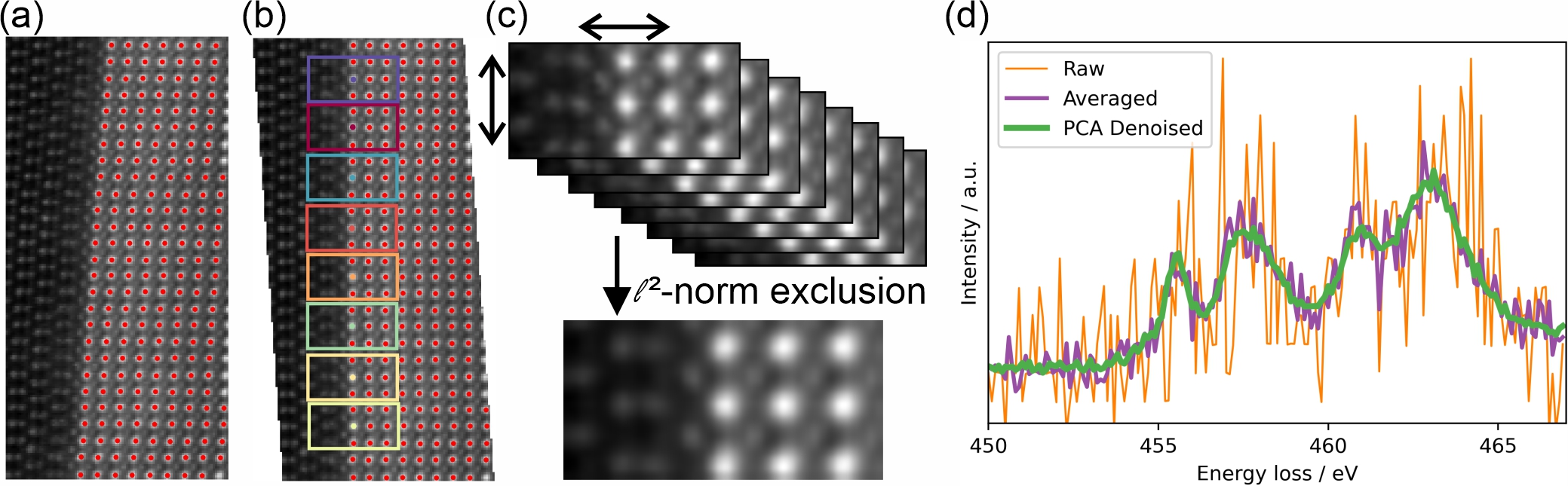}
\caption{Multi-frame approach utilising the periodicity of the crystal. (a) Determined atom positions. (b) Drift-corrected HAADF image with patches, which are cropped from the image. (c) Rigid and non-rigid alignment of the patches. The stack is averaged after it is checked by a L\textsubscript{2}-norm. (d) Comparison of single pixel spectra from the raw and averaged spectrum image. Additionally, a PCA denoising can be applied.}
\label{fig:postprocessing}
\end{figure}

In order to further minimise artefacts due to the alignment and averaging procedure of the single SI patches, the linear specimen drift is corrected. For this purpose, the exact atom positions in the region of interest are determined by the dark field image, utilising the \textit{Python} library \textit{Atomap} (Supplementary~Fig.~\ref{fig:postprocessing}a) \citep{nord2017}. With the knowledge of the theoretical crystal structure and the determined atom positions, the averaged lattice vectors are transformed to the theoretical lattice vectors \citep{wang2018drift}. This rigid transformation is applied to the dark field image and the spectrum image. For cropping smaller frames from the original image, the determined atom positions of the same species are used as centre for each frame (Supplementary~Fig.~\ref{fig:postprocessing}b). These frames are stacked and aligned by a rigid registration based on cross correlation and non-rigid registration based on gradient descent technique to align scan distortions (Supplementary~Fig.~\ref{fig:postprocessing}c) \citep{jones2015}. A maximum shifting length at the rigid registration is set to avoid a jumping to the next atom column. This is critical for example mapping the titanium e\textsubscript{g} states in rutile, because the atoms exhibit nearly no differences in the HAADF reference image, but the shape of the e\textsubscript{g} state is rotated 90° to each other \citep{LOFFLER201726}.

Depending on the quality and sampling rate of the dark field image, the aligning algorithm can fail or the quality of the slice itself can be insufficient, which would degrade the spatial resolution of the averaged image. Therefore, a $\ell^2$--norm is introduced to automatically detect and remove such faulty slices from the stack. The norm is calculated pixelwise between each normalised frame and the normalised averaged image. The slice with the largest $\ell^2$--norm can be eliminated from the stack. The averaged image is then recalculated and the process is iterated until a given number of the frames are removed. This is useful especially at averaging a large number of frames.

Finally, the corresponding spectrum images of the remaining frames are averaged. If the SNR is still too low to map the fine spectral features after averaging, an additional wPCA denoising step should be introduced (Supplementary~Fig.~\ref{fig:postprocessing}d). In practice, the scree plot is used to estimate the number of components, which is assumed to contain all the spectral information for reconstructing the noise-free spectra. Investigating fine spectral features, this number however cannot be determined from the scree plot, increasing the risk of introducing artefacts or removing faint spectral features at reconstruction \citep{lichtert2013,cueva2012}. An interactive selection of number of components is implemented to check if features are only present at a specific number of components, which would indicate an artifact. 

The post-processing code can be downloaded from \url{https://github.com/MichaelO1993/PostProcessing}.

\subsection{Correlative noise}

\begin{figure}[ht]\centering
\includegraphics[width=0.7\linewidth]{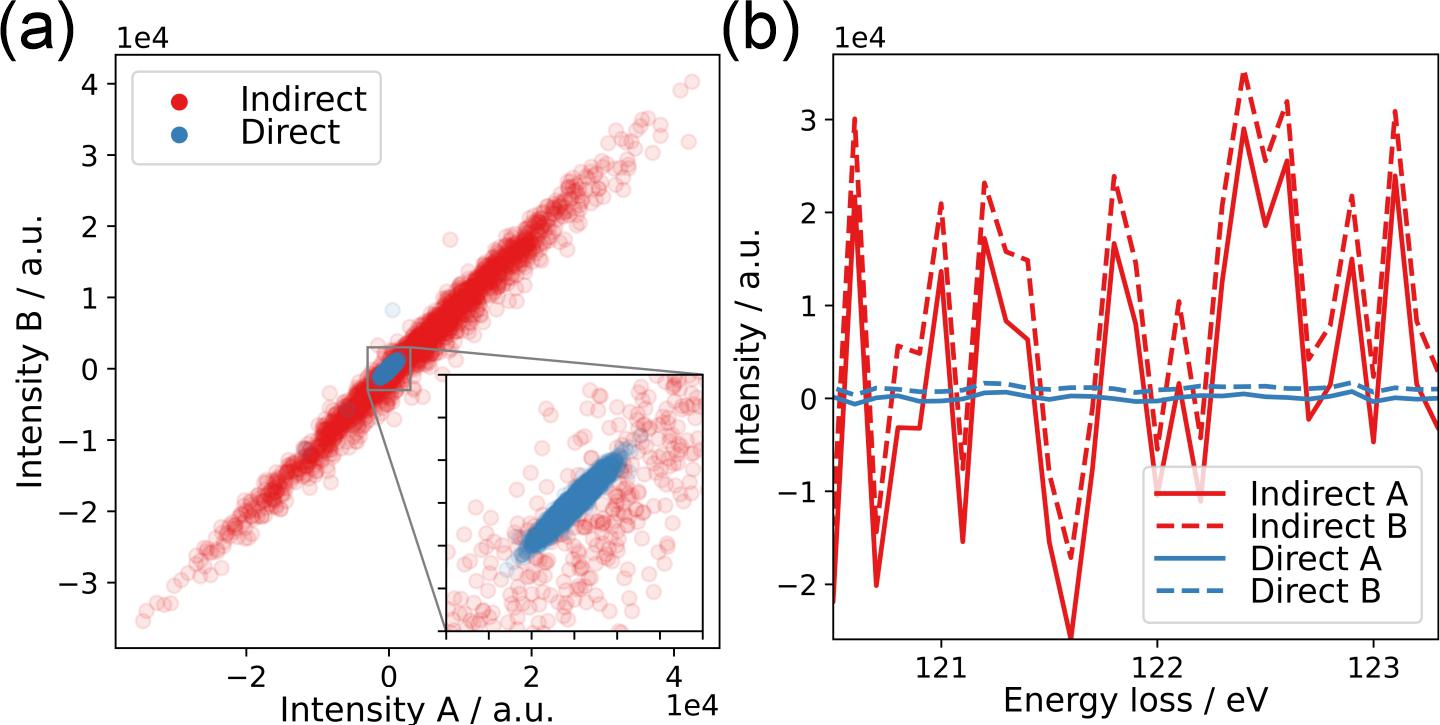}
\caption{Correlative noise of the K2 (blue) and CCD (red). Two spectra, averaged of 22500 spectra of the spectrum image from different regions. (a) Averaged intensities of the region A and B are plotted on the x and y axis as scatter plot. (b) Comparison of the averaged spectra in a small energy loss region. Dotted spectra are shifted upwards for better visualisation of the correlation.}
\label{fig:correlation}
\end{figure}

The direct electron detection camera in counting mode is prone to shot noise only, which should eliminate or reduce the correlative noise in multi-frame imaging caused by the noise of dark and gain references \citep{hart2017}. To verify this statement, STEM-EELS spectrum images with blanked beam are recorded with the direct detection camera K2 and the indirect detection CCD camera UltraScan 1000. Each spectrum image is separated in two areal parts, A and B. The spectra of each area are averaged and plotted as scatter plot against each other (Supplementary~Fig.~\ref{fig:correlation}a), in which every energy loss of the spectra correspond to a point. From these graphs we see that the noise of both cameras is highly correlated. The correlation coefficient of the K2 ($R_2= 0.88$) is only slightly smaller than it is of the CCD ($R_2= 0.98$). Nevertheless, we also note a large difference in the amplitude which means that the correlative noise of the K2 becomes only relevant at averaging a huge number of spectra or investigating very weak features (Supplementary~Fig.~\ref{fig:correlation}b).

\bibliographystyle{unsrtnat}
\bibliography{references}  




